\documentclass[sigchi, screen, anonymous=false]{acmart}
\AtBeginDocument{%
  \providecommand\BibTeX{{%
    \normalfont B\kern-0.5em{\scshape i\kern-0.25em b}\kern-0.8em\TeX}}}
\setcopyright{acmcopyright}
\copyrightyear{2018}
\acmYear{2018}
\acmDOI{10.1145/1122445.1122456}
\usepackage{enumitem,kantlipsum}
\usepackage{hyperref}

\acmConference[Woodstock '18]{Woodstock '18: ACM Symposium on Neural
  Gaze Detection}{June 03--05, 2018}{Woodstock, NY}
\acmBooktitle{Woodstock '18: ACM Symposium on Neural Gaze Detection,
  June 03--05, 2018, Woodstock, NY}
\acmPrice{15.00}
\acmISBN{978-1-4503-XXXX-X/18/06}

\begin{document}

\title{Scrolly2Reel: Retargeting Graphics for Social Media Using Narrative Beats}
\author{Duy K. Nguyen}
\authornote{Equal contribution}
\affiliation{%
  \institution{Columbia University}
  \city{New York City}
  \state{NY}
  \country{USA}
}
\email{hello.duyknguyen@gmail.com}

\author{Jenny Ma}
\authornotemark[1]
\affiliation{%
  \institution{Columbia University}
  \city{New York City}
  \state{NY}
  \country{USA}
}
\email{jenny.ma@columbia.edu}

\author{Pedro Alejandro Perez}
\affiliation{%
  \institution{Columbia University}
  \city{New York City}
  \state{NY}
  \country{USA}
}
\email{pap2153@columbia.edu}

\author{Lydia B. Chilton}
\affiliation{%
  \institution{Columbia University}
  \city{New York City}
  \state{NY}
  \country{USA}
}
\email{chilton@cs.columbia.edu}

\renewcommand{\shortauthors}{Nguyen, Ma et. al.}
\begin{abstract}
Content retargeting is crucial for social media creators. Once great content is created, it is important to reach as broad an audience as possible. This is particularly important in journalism where younger audiences are shifting away from print and towards short-video platforms. Many newspapers already create rich graphics for the web that they want to be able to reuse for social media. One example is scrollytelling sequences or "scrollies" -- immersive articles with graphics like animation, charts, and 3D visualizations that appear as a user scrolls. We present a system that helps transform scrollies into social media videos. By using the scriptwriting concept of narrative beats to extract fundamental storytelling units, we can create videos that are more aligned with narration, and allow for better pacing and stylistic changes. Narrative beats are thus an important primitive to retargeting content that matches the style of a new medium while maintaining the cohesiveness of the original content.
\end{abstract}



\begin{teaserfigure}
    \centering
    \includegraphics[width=\textwidth]{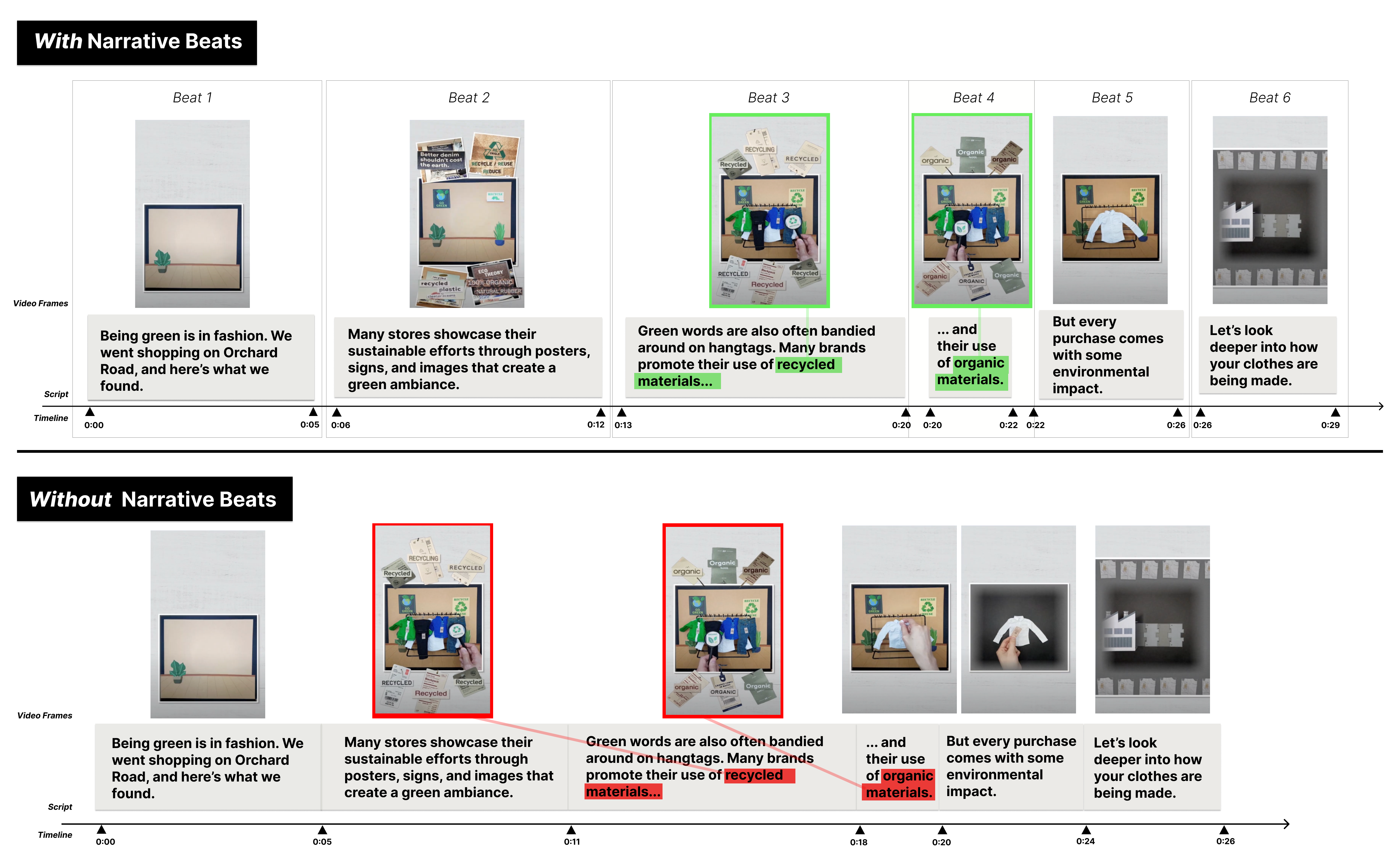}
    \caption{Videos created using Scrolly2Reel utilize narrative beats denoted by text boxes from the scrolly to ensure alignment between the visuals and text. The figure outlines an example using a Straits Time article \cite{Chai2022BeyondTheLabel}. With narrative beats, the video animation is in sync with the audio narration (when the narration says "recycled materials", posters with the words "recycled" appear, and when "organic materials" is said, posters saying "organic" also appear), creating a satisfying viewing experience. In the video without narrative beats, the visuals are not controlled with regard to the scrolly's beats ("recycled" and "organic" posters are shown before the audio reference), resulting in misalignment and confusion.}
    \label{fig:alignment}
\end{teaserfigure}

\maketitle

\footnote{Github Repository: \href{https://github.com/duynguyen158/Scrolly2Reel}{https://github.com/duynguyen158/Scrolly2Reel}}
\footnote{Project Page: \url{https://jennygzma.github.io/\#/scrolly2reel}}

\section{Introduction}

Content retargeting is crucial aspect of generating social media videos. Once great content is created, it is important to transform it so it can reach many different audiences on many different platforms. This includes changing its format (turning text into video), changing its length (making a long videos into a short video), and changing its style (appealing to a younger or older audience). However, content retargeting is rarely as simple as cropping an image to a new size or summarizing an article to a different length. It requires understanding the original medium's style and transforming it to the target's style, while keeping its message coherent. Although content retargeting is a general problem affecting all content creators, it is particularly important in journalism where publishers seek to inform and engage the general population in the most important issues of the day. 

Journalism masters the art of telling stories. As media evolves, storytelling evolves. With the rise of digital journalism came immersive, media-rich stories that use the ability to scroll a page as a mechanism for users to unfold the graphics and stories at their own pace. These stories are called scrollytelling stories (or ``scrollies''). The New York Times' 2012 digital article called Snow Fall \cite{SnowFall} is credited as the first example of this scrollytelling technique. It won a Pulitzer Prize \cite{pulitzer2013} for its narration of a tragic avalanche in Washington State that illustrates a moment-by-moment account of the skier's paths as the avalanche tore down the mountain. 
This article's combination of self-paced 3D graphics guided by minimal text helped usher in a new era of multimedia journalism and digital storytelling~\cite{Bahr2022SnowFall,Greenfield2012SnowFall}.


Storytelling is evolving again with the rise of social media. People under the age of 25 worldwide turn more and more to social media for news~\cite{newman_news_tt_reuters_oxford}. The current trend is reels - short-form videos spread over platforms like Instagram and TikTok. Journalists are currently exploring this trend as a mechanism to deliver the news, with many notable successes. 
For a news reel to be successful, it has to capture and sustain attention and  communicate its story quickly and clearly to prevent users from scrolling past it. ~\cite{savolainen2022infotainment}. 
Stories meant for print media target readers with longer attention spans and more investment in the story. Retargeting this content for social media requires a significant restructuring of the content and reimagining of the narrative.



Content retargeting is time-consuming and thus there has been much interest in using computational techniques, deep learning, and generative AI to assist in the process. This includes examples such as using specially trained deep neural networks to reshape images from a web banner to a social media title \cite{contentatscale2024, vertexai2024, berent2024}, using generative AI to extract the most emotional clips from videos to create "teaser" videos \cite{wang2024podreels}, and using generative AI to turn journalism print articles into scripts and storyboards for reels for creators to act out \cite{reelframer}. Generative AI is particularly powerful as a technology for content retargeting because it allows for flexible extraction and rearrangement of material. However, the extraction and rearrangement must still maintain a coherent narrative, and thus, it is useful to have structures and concepts to help with content transformation.





To explore transforming digital assets for news reels, we propose using \textbf{narrative beats} 
to retarget from one type of content to another.
In screenwriting and other forms of storytelling, narrative beats (or just \textit{beats}) are small moments that move the story forward. They are a fundamental unit of storytelling ~\cite{Perelman2023StoryBeat}.
Beats are not explicitly marked in most media. Instead, they are a conceptual grouping of information, similar to line numbers in a script or timestamps in a video. Beats can be inferred from the original medium and underpin the transformation of content to a new medium.

By identifying the beats within a scrolly, we enable three ways of reshaping the narrative: 
\begin{itemize}
\item \textbf{Align the visuals with the narration.} When creating the video, the visuals should be scrolled at a speed such that relevant text is only read while the relevant graphics are showing. This makes the video more coherent. The importance of alignment can be seen in the video example of Figure \ref{fig:alignment}.
\item \textbf{Control the pacing.} Social media videos are typically faster-paced than text. By shortening the text within a beat, the video's narration has a faster and more engaging pace. 
\item \textbf{Alter the style.} Spoken text has a different style than written text. Within a beat, rewrite the text to be more natural and easier to understand when spoken. 
\end{itemize}

\begin{figure} 
    \centering
    \includegraphics[width=\columnwidth]{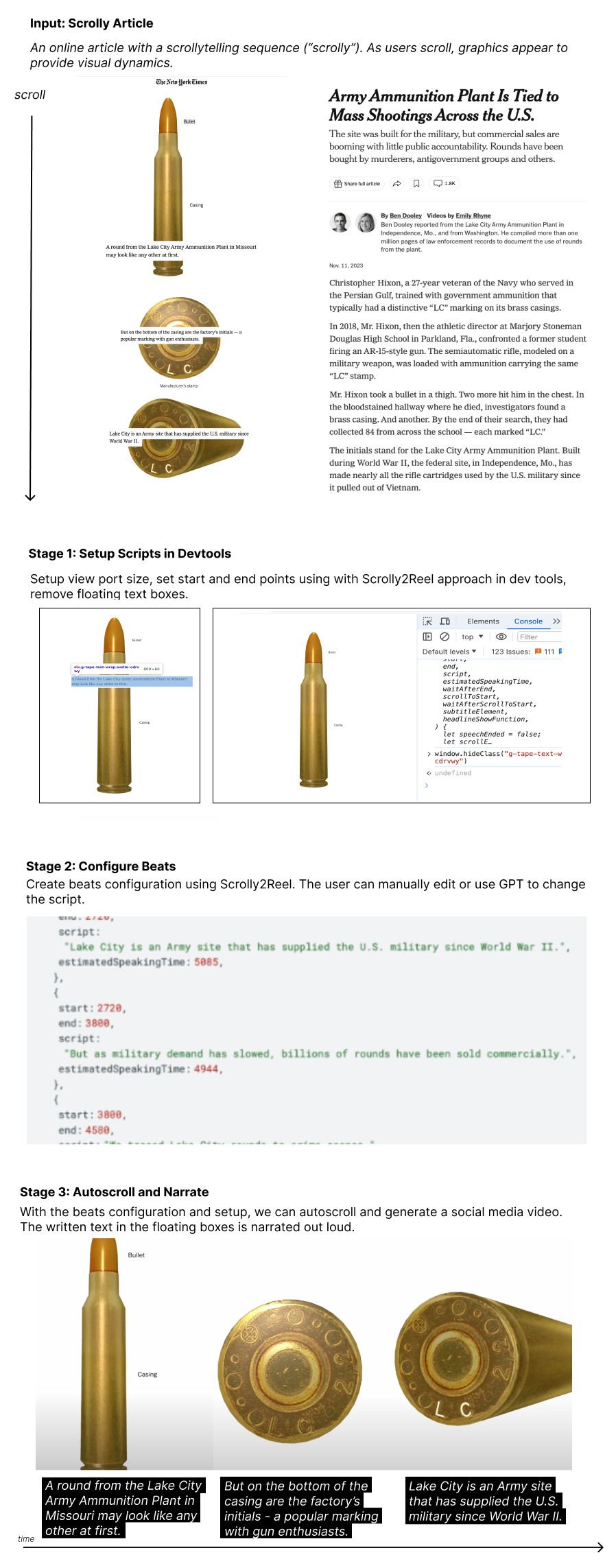}
    \caption{Scrolly2Reel provides an approach to convert scrollies in articles to a social media video with beat alignment. The example above shows a New York Times scrolly ~\cite{Dooley2023AmmunitionPlant}}.
    \label{fig:system}
\end{figure}

 We introduce Scrolly2Reel, which allows users to capture videos based on a scrolly. It is implemented as a script that runs in Developer Tools of Chrome Browser. A step-by-step walkthrough can be seen in Figure \ref{fig:system}. For a scrolly, we identify beats by finding the text boxes floating above the images, as seen in Stage 1 of Figure \ref{fig:system}, as these floating text boxes naturally move the scrolly-telling story forward.
We extract the placement of the text boxes within the page and use the page offset to determine the visuals that correspond to it. We can then automatically scroll through each beat while narrating the text boxes, and proceed to the next beat only when the speaking has concluded, keeping the narration aligned with the visuals. To speed up the pace and make the text suitable for narration, the text of each beat is shortened and simplified using a LLM prompt with instructions and few-shot examples. 

In an expert evaluation of 17 scrollies,
we find that automated videos with beat alignment and text-shortening within each beat are most preferred. Videos using the raw article text as the script are too long, leading to a lack of immersion in the story. Videos without beat alignment are occasionally acceptable if the text naturally aligns with the visual segments. However, in many cases, such misalignments lead to confusion and impede the narrative of the story. We conclude that in retargeting content, discovering the beats that represent a story at its fundamental level is computationally feasible, which allows for successful retargeting and enables stories to reach new audiences.

\section{Related Works}


\subsection{Scrollytelling and Digital Media in Journalism}

Media is a powerful narrative tool in journalism. Even in traditional journalism forms such as print and television, images and videos are key to capturing, engaging, and informing audiences. Digital media has only expanded the tools by which journalists can tell stories with interactive visualization ~\cite{2010-narrative}. This includes interactive graphs, a photo gallery for users to click through, and more recently scrollytelling, a digital news design in which a narrative with both text and visuals is told at the user's pace as they scroll.

Scrollies often use rich graphics to help tell complex stories~\cite{8564193}. Often, this includes explaining how things happen: how a building collapsed in Miami~\cite{MiamiHerald2021HouseOfCards}, how COVID spreads through the subway~\cite{NYT2020SubwayVirus}, and how we know that women of color are more likely to die in childbirth than white women~\cite{NYT2021MaternalHealth}. These stories are expensive to produce so they are not typically used for breaking news, but are often used for complex analyses or investigative stories, which are still relevant months, days, or years after the original story breaks~\cite{8564193}. They have high production value as multiple tools have emerged to help journalists and graphic designers produce scrollies~\cite{Moerth2022ScrollyVis,dataparticles,2018-idyll}. 
 
\subsection{Narrative and Narrative Visualization}
Narrative beats are small moments that move the story forward ~\cite{Perelman2023StoryBeat}. They are a fundamental unit of storytelling which are extensively used in video-based media. For example, in modern television shows each beat typically lasts  two minutes; this is how often the audience needs to see something new to hold their attention~\cite{Newman2006BeatsArcs}. In different media, beats can have different lengths and formats, but the goal is always the same - to sustain the audience’s attention at a regular pace. When creating video content, understanding the beat structure and pacing is key to audience engagement.


Narrative is traditionally thought of as part of a written or spoken medium, but it also has an analog in visualization. Studies of visual narratives break down the “tactics” for visual narratives into three types: 1) visual structuring (how the overall story is laid out), 2) highlighting (what information is emphasized), and 3) transition guidance (how to move between “visual scenes” without disorienting the viewer ~\cite{2010-narrative}. When retargeting scrollies into videos, the visual structuring and highlighting can stay the same; but we need to decide where the new transitions will be. Particularly, the visuals need to align with the spoken text - visuals should not transition until the relevant text is spoken.





\subsection{Content Retargeting}
There has been much research within the realm of content retargeting and using generative AI to assist in the process. Retargeting content is rarely as simple as cropping an image to a new size or summarizing an article to a different length. It requires understanding the original medium's style, extracting the underlying structure and content, and then mapping that into the structures and styles of the new medium. Understanding the primitive representations of the narrative of both domains is essential to successful retargeting.

As young people turn more and more to social media for news, stories meant for print media must be reframed for those with shorter attention spans, while still retaining the original narrative quality. 
This means subjecting the generated output under stylistic constraints such as brevity and formality of language, on which prior effort has been made~\cite{sawicki2022training}. Our work in retargeting content for new and younger audiences is preceded by many other examples, including reframing complex scientific topics into Tweetorials on what was previously Twitter \cite{tweetorial_hook}, transforming longer topics such as financial learning to entertaining experiences \cite{rasco2020fincraft}, and transforming journalism print articles into scripts and storyboards for reels and creators to act out. \cite{reelframer}.




\subsection{News in Social Media Videos}

News reels are an emerging trend in journalism to help news outlets reach younger audiences who get their news almost exclusively from social media and, increasingly, from short-form videos on TikTok and Instagram~\cite{newman_news_tt_reuters_oxford}.
A 2022 report from the Reuters Institute found that 49\% of top news publishers worldwide are regularly publishing content on TikTok, most of whom started in 2021 or later~\cite{newman_digital_news_2022}. 
Additionally, they found 15\% of 18-24-year-olds already use TikTok for news~\cite{newman_digital_news_2022}; there is an appetite for news content in this new format.

In interviews conducted by the Reuters Institute in 2022 with news reel creators~\cite{savolainen2022infotainment}, many different approaches to creating reels have been found, but a key challenge is to balance between information and entertainment~\cite{savolainen2022infotainment}. 
Information value is important to news, but entertainment is also a key component because
if a video does not grab and hold an audience's attention, they may scroll past it. 
Thus, successful news reels are fast-paced: audio comes quickly and visuals are always moving and transitioning quickly ~\cite{reelframer}.


\section{Scrolly2Reel}
We explore how to computationally create social media reels using the visual and textual information scrollies contain. We use a JavaScript script meant to be run in the developer console of Google Chrome. At a high level, it identifies the text boxes representing beats in the scrolly, hides them from view, and turns them into an audio script to narrate. It then automatically scrolls through the scrolly while narrating, during which the user records their computer screen and audio to create the product reel.

Our approach targets news graphics editors who can send the produced short video to a social media editor for downstream editing. We assume typical news graphics editor’s qualifications: basic knowledge of the browser developer tools, namely the HTML element inspection tool, the developer console, and front-end programming languages (HTML, CSS, JavaScript), and experience working with chatbots such as ChatGPT.


\subsection{Stage 1. Set up scripts in DevTools}
\subsubsection{Set up the viewport}

Once the user has opened the scrolly in Chrome, they can open Chrome developer tools and change the viewport to a mobile size that conforms to a 9:16 aspect ratio, a typical aspect ratio for reels (commonly 540 x 960 pixels). Due to its responsive design, the scrolly's appearance changes across different viewport sizes to ensure usability across different device types and renders well after resizing.
\subsubsection{Identify the start and end points for the autoscroll process.}

The user needs to identify where to start scrolling and where to stop. These two points also help identify and hide the text boxes in the next step. We offer a function that helps; the user can scroll to their desired start position and run the function to grab its pixel value. They can then scroll to their desired end position and run our function again to get its pixel value. 
\subsubsection{Identify text boxes and hide them from view.}

The user needs to identify the text boxes in the scrolly for three reasons: 1) to use their position to segment the story into beats, 2) to use the text they contain as the basis of the voice narration, and 3) to remove them while the user records the video of their screen for the reel. They can either identify the class name of the text box and hide it using CSS, or use a function we provide to identify the element using the text content of the box and remove all of the text boxes from the page. 

\subsection{Stage 2. Configure Beats}
\subsubsection{Get Beats Configuration}

With the viewport size set and the start, end, and beats identified, the user needs to set up their beats configuration to automatically scroll the article and narrate. To do so while keeping synchronicity with the beats, we represent the scrolly as a list of beats. An example of this configuration can be seen in Figure 2, Stage 2.

\subsubsection{(Optional) Shorten and/or concatenate the beats.}

The user can edit the beats representation obtained to derive new representations to make the video more immersive. For example, to improve pacing, they can shorten the beat texts using GPT-4 as a way to reduce the estimated speaking time of a beat, thereby making the narration faster. We provide a system prompt where a user can input the original beats representation and have the same list of beats, each with its text shortened, returned by GPT. Additionally, the user could concatenate beats, or manually modify the script within each beat to make the content more engaging.
\begin{figure*}[!htb]
    \centering
    \includegraphics[width=\textwidth]{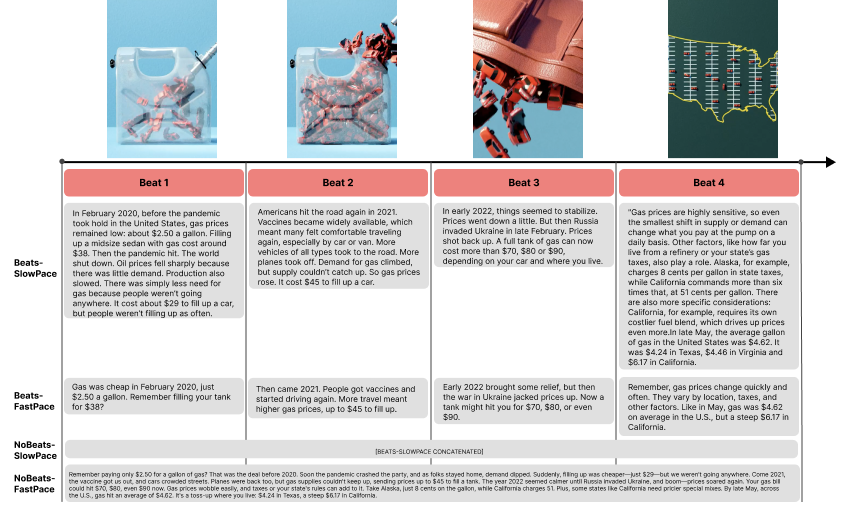}
    \caption{Beat sequences with scripts and keyframes across different reel versions, taken from a Washington Post scrolly~\cite{WP2022GasPrices}}
\end{figure*}



\subsection{Stage 3. Autoscroll and Narrate}
With the beats configuration, the user can use our function to automatically scroll the article and narrate with text that aligns with the visual elements of the scrolly. Our function utilizes a linear scrolling process, which keeps the scrolling speed constant within each beat. The scrolling speed (pixels per second) is calculated using the start and end position (in pixels) and the estimated time it takes to speak (in seconds).
\begin{equation}
\text{scrolling speed} = \frac{\text{end position} - \text{start position}}{\text{estimated speaking time}}
\end{equation}

The script abstracts away this calculation through the use of the animate function in the jQuery JavaScript package~\cite{jQueryAnimate}. The script iterates through each beat, scrolling through and narrating the script using the browser’s synthetic voice. To create a reel, the user simply needs to record their screen and audio after calling the function. This reel can then be passed into an editing software such as CapCut to be further refined, for example, by adding a subtle background track.
\section{Evaluation}
To evaluate the key principles for transforming scrollies into reels, we test two variables, alignment and pace. To test the alignment between the visuals and narration, we have videos created by retargeting scrollies' narrative beats into the generated videos, and videos created without regard to narrative beats. Our beat-driven videos, called ``Beats", are created by the system approach we defined above, and generated to synchronize audio and visual elements. Videos without beats, termed ``NoBeats," are created by concatenating all beats into one, removing intentional alignment.

To test pace, we have fast-paced videos and slow-paced videos. Our slow-paced videos, termed ``SlowPace", use scripts automatically generated from the article that retain the original article text. For fast-paced videos, called ``FastPace", we paraphrase the SlowPace narrative script using a GPT prompt, and our scripts in Scrolly2Reel adjusts animation speed to match the shortened script automatically. 

\subsection{Data}
We picked a total of 17 scrollies from 17 different articles that won Gold and Silver prizes in the digital category in Best of News Design competitions held by the Society of News Design between 2021 and 2023~\cite{SND42MedalWinnersDigital,SND43DigitalResults,SND44Winners}. We created four versions of each reel using the conditions described above for a total of 68 videos\footnote{\href{https://www.youtube.com/playlist?list=PL6FWuSZeaOxdWztwsoiLIAUijqL1cgCx1}{Sample Videos}}

\begin{enumerate}
    \item Beats-SlowPace: a video with beats and no GPT shortening.
    \item Beats-FastPace: a video with beats and GPT shortening. 
    \item NoBeats-SlowPace: a video with no beats and no GPT shortening.
    \item NoBeats-FastPace: a video with no beats and GPT shortening.
\end{enumerate}

\subsection{Participants and Procedure}
We evaluated our reels across four dimensions: 1) attention, 2) immersion, 3) alignment, and 4) pacing. Attention measures whether or not the reel captures the viewer's attention within the first couple seconds. In a social media context, it determines whether users would continue watching, or skip the reel. Immersion refers to the reel’s ability to maintain the viewer’s attention throughout the animation. Alignment assesses how well the speech in the reel matches the visual elements of the story, as it’s essential to ensure synchronicity between audio and visuals. Pacing denotes the speed of the narrative in the reel. A well-paced reel is easy to follow and keeps the viewer engaged. 

We recruited two annotators with experience in both journalism and social media from a local university. Both were undergraduate students with 3+ years of journalism experience (average age 21, 1 male, 1 female). They independently rated the 68 videos for attention, immersion, alignment, and pacing on a scale of 1 (bad) to 5 (good). Videos were presented in random order without indication of the conditions that generated the videos. Annotators were introduced to the task through a teleconference call, shown the examples, calibrated the videos with the of the rubric, and compensated for their time.


We hypothesize that videos in the Beats-FastPace condition will outperform the other versions across all dimensions: attention, immersion, alignment, and pacing. Specifically,

\textbf{H1)} \textit{(a)} Videos that use GPT to shorten the text (FastPace) will be overall preferred to videos generated with original text (SlowPace); \textit{(b)} Videos that use GPT to shorten the text (FastPace) will score higher on the pacing dimension.

\textbf{H2)} \textit{(a)} Videos with beat alignment (Beats) will be overall preferred to videos with no beat alignment (NoBeats); \textit{(b)} Videos with beat alignment (Beats) will score higher on the alignment dimension.


\subsection{Results}
\subsubsection{Quantitative Results}

\begin{table*}[ht!]
  \centering
  \begin{tabular}{lccccc}
    & Attention & Immersion & Alignment & Pacing & Overall \\
    \hline
    Beats-SlowPace & 3.59 & 2.88 & 3.21 & 3.26 &	3.23\\
    Beats-FastPace & 3.74 & 3.44 & \textbf{3.59} & \textbf{4.09} & \textbf{3.71} \\
    NoBeats-SlowPace & \textbf{3.88} & 3.15	& 3	& 3.21 & 3.31\\
    NoBeats-FastPace & 3.85 & \textbf{3.47}	& 2.91 & 3.97 & 3.55 \\
    \hline
  \end{tabular}
  \caption{Attention, Immersion, Alignment, Pacing for Different Variants. Bold indicates the highest number in the column.}
  \label{tab:interrater-reliability}
\end{table*}
 
We first averaged the ratings from the two annotators to see which conditions performed best across the 4 dimensions. We also averaged the score across all dimensions to get an overall score. The results can be seen in Table 1 (the highest score in each column is denoted in bold). The Beats-FastPace videos were hypothesized to be the best, and it has the highest overall score. (3.71 out of 5). It also has the highest score for alignment and pacing, and while it is not the highest score for attention and immersion, it’s close to the top.

To test for significant differences, we ran ANOVA tests for the overall score and found that there were indeed differences between the conditions at the p\textless0.05 level (p = 0.0376). However, to determine which conditions were the best, a Tukey’s HSD test showed that only one pair was statistically significant - Beats-FastPace was statistically different from Beats-SlowPace. Beats-FastPace was not statistically higher than the other conditions. \textbf{This shows partial support for H1a}, that FastPace videos will be preferred to SlowPace videos. Unfortunately, \textbf{there were not statistically siginificant results to prove H2a}, that Beats videos will be overall preferred to NoBeats videos. 

We moved on to conduct analysis for each dimension: pacing, attention, immersion, and alignment. For the attention and immersion dimensions, the ANOVA tests showed there were no significant difference across means, given its inherently subjective nature. We moved on to do post-hoc comparisons using Tukey’s HSD test. While no discernible variance was observed in attention and immersion dimensions, we found statistically significant differences along the pacing and alignment dimensions. 

The ANOVA tests for the pacing dimension show that there was a significant difference between the conditions at the p\textless0.01 level (p=0.00357). Additionally, a Tukey’s HSD test showed that there were statistically significant differences, demonstrating that the FastPace conditions all had better pacing than the SlowPace conditions. Beats-FastPace had a significantly better pace rating than both Beats-SlowPace (p=0.0317) and NoBeats-SlowPace (p=0.0186) NoBeats-FastPace had a significantly better pace rating than both Beats-SlowPace (p=0.0842) and NoBeats-SlowPace (p=0.0526). \textbf{This shows full support for H1b.} 

The ANOVA tests for the alignment dimensions illustrated a difference in alignment with p\textless0.1 (p=0.0667). Both overall beat alignment scores were higher than no beat alignment, but were only significant for FastPace videos. Beats-FastPace videos displayed statistically significantly better alignment compared to NoBeats-FastPaced videos at the p=0.10 level (p=0.067). However, Beats-SlowPace videos did not show statistically significant better alignment than Beats-FastPace, \textbf{displaying partial support for H2b}. 

Although the quantitative analysis does not support the idea of beats, logically it is clear that there should be some cases where having beats aligned to the video must be better than having beats that are not aligned to the video. Thus, we conduct a qualitative analysis to understand the role of beat alignment in the quality of a reel.

\subsubsection{Qualitative Results}

In our qualitative analysis, we look at the importance of beat alignment. To do this, we analyzed the 17 videos in the dataset to see how many beat alignments were better, same, or worse, when fixing their pace condition (Beats-FastPace vs NoBeats-FastPace, and Beats-SlowPace vs NoBeats-SlowPace). We defined ``better'' as videos where the Beats video was rated 0.5 points higher than the NoBeats video. We defined ``worse'' as videos where the Beats video rated 0.5 points lower than the NoBeats video; we defined videos where the Beats video were less than 0.5 points greater than the NoBeat video as ``same." 

\begin{table}[ht!]
  \centering
  \begin{tabular}{lcc}
    & FastPace & SlowPace \\
    \hline
    Better than NoBeat by $\geq$ 0.5 & 10 & 9\\
    Same as NoBeat ($\leq$ 0.5) & 3 & 4 \\
    Worse than NoBeat by $\geq$ 0.5 & 4 & 4\\
    \hline
  \end{tabular}
  \caption{Comparison of Beat vs NoBeat in Alignment}
  \label{tab:interrater-reliability}
\end{table}
In FastPace videos, 10 of 17 videos with beats were rated better than with no beats, three were rated the same, and four were rated as worse. It is unsurprising why 10 videos were rated better with beats. Without the beats, some visuals become unaligned with the text. The audio is referring to things that are no longer on screen - or worse - have not shown up on screen yet. This is disorienting to viewers.

For the three videos where beat alignment was rated the same for both NoBeats and Beats, we noted that the audio and visuals coincidentally aligned with the video without discernible errors for the NoBeats videos. These videos were all less than 60 seconds; even without beats, in short videos, it is possible for alignment to happen naturally, while for longer videos, it’s easier for visual and audio elements in to diverge. Additionally, for some of these videos, we observed that the audio is not as closely tied to the visuals, meaning that the alignment is less important.

Surprisingly, four videos actually performed worse with beat alignment that without it. In reviewing them, we saw that sometimes the NoBeat videos coincidentally mentioned something in the script was directly mirrored in the animation. A comparison with our beat condition videos revealed a slight delay between the script and the on-screen presentation. This is due to our alignment algorithm - because we start the linear autoscroll of a beat right when the beat’s narration starts, what is presented on the screen can slightly lag behind what is said in the narration. Conversely, in the NoBeat condition, for coincidences when the script immediately matched the animation, it created moments of excitement, capturing viewers' attention and greatly improving alignment scores. These videos were also approximately 30 seconds in length; alignment tends to occur more naturally in shorter videos.

We found similar patterns in the SlowPace videos. For all the Beat-SlowPace videos that did not outperform their NoBeat-SlowPace counterparts, the videos were 30-60 seconds in length, short videos where alignment occurs more naturally. They also had coincidental moments where the script matched the animation, resulting in high alignment scores. 

We conclude that not all scrollies require visual and script alignment, and in those cases, utilizing the narrative beats to retarget the content is less crucial. We also observed happy coincidences of alignment without beats when the videos were short. Ultimately, however, for scrollies where the text is heavily tied to the graphics, in its video form, moments where the script matched animation were exciting and engaging, illustrating that beats are necessary to preserve alignment and narrative cohesion in retargeting content. 
\section{Discussion}
In this paper, we introduce narrative beats as a  
concept to help with content retargeting. Using narrative beats as a fundamental representation of the original content helps restructure it into new forms that maintain the coherency of the narrative but can change the modality and stylistic elements like pacing to fit a new format. We show that using beats helps transform scrollytelling print articles into social media videos. By extracting beats from the print article, we can 1) \textbf{transform the medium} by extracting the text from each beat, narrate it, and film a video while the article autoscrolls 2) \textbf{align the visuals with the narration} by adjusting the scrolling speed of each beat to match the length of the narration of the beat text, 3) \textbf{Adjust the pace of the video} by shortening the narrated text within each beat, and  4) \textbf{Change the style of the narration} within each beat to better suit to listening (rather than reading). Together, these transformations can reuse the original story and graphics while creating an immersive social media experience that grabs and holds attention.

In our quantitative analysis, when retargeting scrollies to reels, videos with narrative beat alignment and text-shortening performed the best overall. Additionally, we found that shortening the text within beats is crucial for faster and more engaging pacing. In our qualitative analysis, we took a deeper look at our videos generated and found that videos where the visuals directly referenced the narration produced higher scores. In some cases the alignment occurred coincidentally; however, using narrative beats in retargeting scrollies better facilitates this magical alignment so that it is not accidental but concrete.


Narrative beats are a fundamental unit of storytelling, and they exist across all storytelling media. Print stories have paragraphs, comics have panels, slide shows have slides (and sometimes animations on slides, social media reels have cuts, video games have game states). Some beats are easier to extract that others, but beats almost always exist.  Although in this paper, we extracted beats using a set of code functions that we wrote by hand, generally, AI can more flexibly extract information and break it up into sections (or beats), even with fuzzy definitions. The flexibility of AI is also powerful to make a transformation of that material, such as summarizing text or making dry explanatory text into a dialog between characters \cite{reelframer}. Content retargeting is crucial for reaching multiple audiences on social media. More creation tools that allow for the flexible reuse of content would be hugely valuable.



\section{Future Work and Limitations}
Our study relied on trained raters, whereas social media content targets younger audiences. Future investigations could use less-trained raters to get diverse opinions and achieve more statistically significant data. Our analysis was also limited to 17 scrollies from award-winning articles. We could expand our dataset to include more interesting analyses to ensure that our approach can be applied in various contexts. 



We can additionally fine-tune our alignment within the scrolly using beats. Concatenating beats into more logical groupings to fit a video framework can be explored. Additionally, as we saw in the qualitative analysis, even though beats encourage alignment, sometimes lagging still occured due to our linear scrolling algorithm and what is presented on the screen is slightly behind what is said in the script. We could explore further precision of alignment within beats when it is especially imperative - for example, when for half a second the visual has to sync up with particular word in the script. 


\section{Conclusion}
This paper explores narrative beats and pacing in computationally generated videos derived from scrollies. Our approach introduced a novel way to grab beats and synchronize them with scrollies, while also condensing the article text for a paraphrased audio script. By synchronizing the audio narrative with the visual elements of the original scrollies and paraphrasing the script of the videos, we produce higher quality, faster-paced animations that are visually interesting while retaining the original content of the scrolly. A quantitative evaluation showed that GPT-shortened videos resulted in significantly higher scores in pace. A qualitative analysis showed that for videos where alignment is crucial, beat synchronization is the key. Our emphasis on beats and pacing acts as an essential component for alignment and narrative interest when retargeting content for different creative mediums.

\bibliographystyle{ACM-Reference-Format}
\bibliography{bib}

\appendix
\onecolumn

\section{Appendix}

\subsection{GPT-Prompt for FastPace videos in Evaluation}
This prompt was pasted into OpenAI's playground as Instructions for the assistant section. 

\subsubsection{Model:} gpt4-1106-preview

\subsubsection{Prompt:} You are an assistant for a news editor. Your task is to paraphrase written news text to be spoken word.  Your input is a list of snippets - your output list needs to be the same length, summarizing 1:1 according to corresponding input snippets. Simplify the language and use imperative language. Ensure cohesion in the new text. Keep references to the actual news article - ie: demonstrative pronouns like “this”. You may draw on information from other snippets to inform paraphrasing on a singular output snippet. Please keep sentences 5-10 words in length, and at most one or per sentence. You can have multiple sentences per output snippet. Try to make the first output snippet as catchy as possible. 

\textbf{Example input and output:}

\textit{Input:}
["A round from the Lake City Army Ammunition Plant in Missouri may look like any other at first. But on the bottom of the casing are the factory’s initials — a popular marking with gun enthusiasts.",
"Lake City is an Army site that has supplied the U.S. military since World War II.",
"But as military demand has slowed, billions of rounds have been sold commercially." "We traced Lake City rounds to crime scenes.",
"For instance, 84 Lake City rounds (of 147 total) were fired in the Parkland school shooting."
]

\textit{Output:}
["Look at this bullet and the letters on its casing.",
"L.C. That's Lake City Army Ammunition Plant. It's an Army site that makes bullets for the military.",
"But it has also sold these bullets to the market. And civilians have used them in mass shootings.",
"In the Parkland school shooting, more than half of the bullets fired were Lake City bullets."
]

\begin{figure}[!htb]
    \centering
    \includegraphics[width=0.8\linewidth]{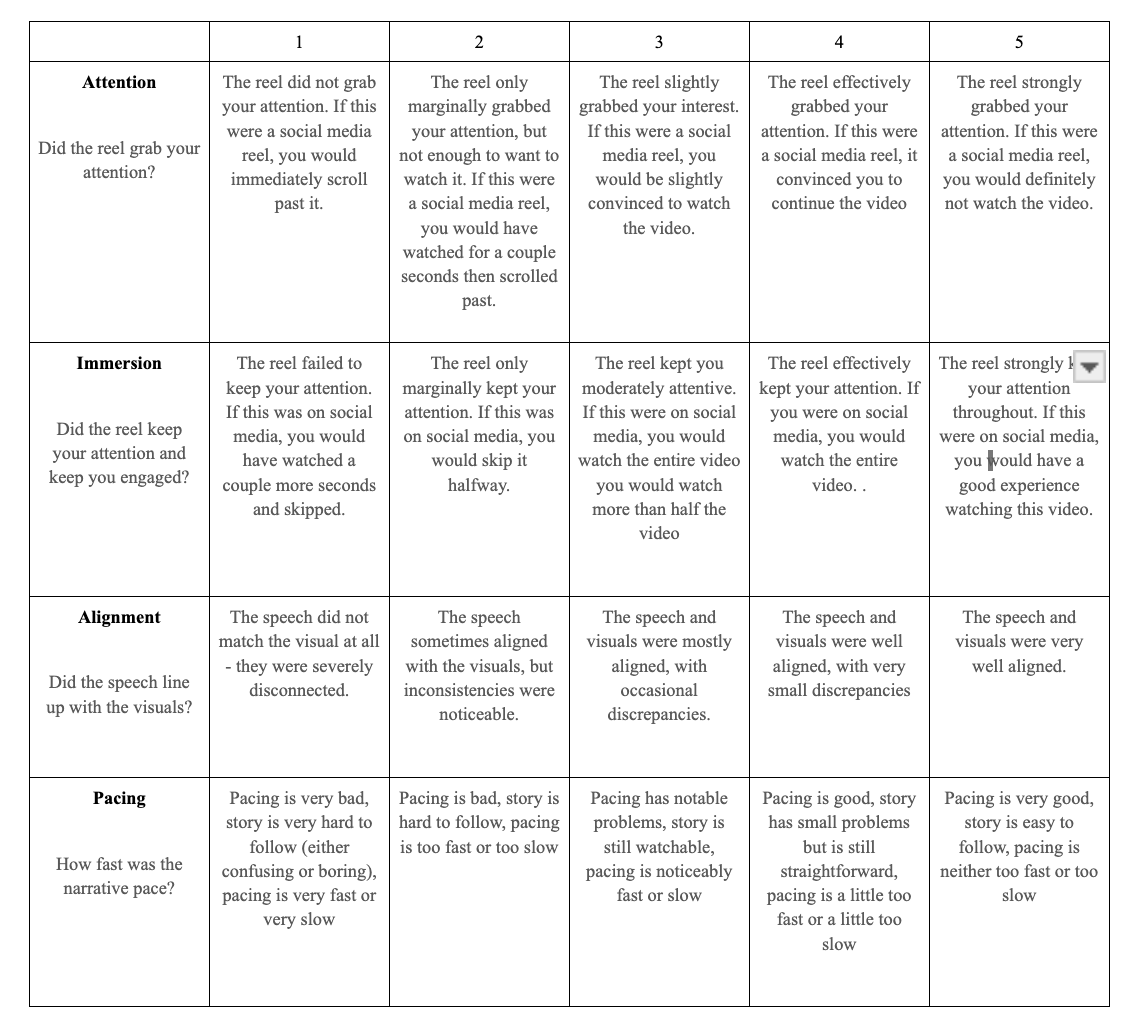}
    \caption{A five-point scale rubric for annotators in the evaluation}
    \label{fig:enter-label}
\end{figure}

\makeatletter
\makeatother
\end{document}